\documentclass[seceq]{ptptex}
\usepackage{graphicx}
\title{Two Pion Production from Nuclei}

\author{
Susan \textsc{Schadmand}$^{1,}$\footnote{ e-mail address:
s.schadmand@fz-juelich.de}
}

\inst{
$^1$ Institut f\"ur Kernphysik, Forschungszentrum J\"ulich,
D-52425 J\"ulich, Germany
}

\abst{
Differences in the photoproduction of mesons on the free proton
and on nuclei are expected to reveal changes in the properties of
hadrons.
Double pion photoproduction from nuclei has been used
to investigate the in medium modification of meson-meson interactions
with the TAPS photon spectrometer at MAMI-B. Recent developments are reported.

Hadron-hadron interactions and meson production and decays involving
light nuclei are also studied in hadron induced experiments.
In particular, the physics program of the new WASA-at-COSY facility includes
double pion production from light nuclei.
Here, the efforts focus on the ABC effect, the puzzling low-mass enhancement
in the $\pi\pi$ invariant mass spectrum, first observed in inclusive measurements
of two-pion production in nuclear collisions.
}

\begin{document}

\maketitle


\section{Photoproduction of Pion Pairs from Nuclei}

The study of in-medium properties of mesons and nucleon resonances
carries the promise to find signatures for partial chiral symmetry
restoration at finite baryon density and temperature.
A particularly interesting case is the mass split between
the $J^{\pi}$=0$^{-}$ pion and the $J^{\pi}$=0$^{+}$ $\sigma$-meson.
The naive assumption that the two masses should become degenerate
in the chiral limit is supported by model calculations
\cite{Bernard:1987im}.
However, the very nature of the $\sigma$ meson is a matter of debate.
The review of particle properties \cite{Yao:2006px} lists
the $f_{0}$(600) with a mass range from 400 - 1200 MeV and a
full width between 600 MeV and 1000 MeV.
Recently, precise predictions for mass and width from
dispersion relations have been derived in \cite{Caprini:2005zr}.
The $\sigma$ meson is treated as a pure $q\bar{q}$ (quasi)bound state
\cite{Bernard:1987im,Hatsuda:1999kd,Aouissat:1999vx},
and as a correlated $\pi\pi$ pair in a $I=0$, $J^{\pi}=0^+$
state \cite{Chiang:1997di,Roca:2002vd,Chanfray:2004vb}.
In all cases, a strong coupling to scalar-isoscalar pion pairs and
a significant in-medium modification of the invariant mass
distribution of the pion pairs is predicted.
This is either due to the in-medium spectral function of the $\sigma$ meson
\cite{Hatsuda:1999kd} or the in-medium modification of the pion-pion interaction
\cite{Roca:2002vd} due to coupling to nucleon - hole, $\Delta$ - hole and
$N^{\star}$ - hole states.
The predicted effect is a downward shift of the strength in the invariant
mass distributions of scalar, isoscalar pion pairs in nuclear matter.

First experimental evidence had been reported by the CHAOS collaboration 
from the measurement of pion induced double pion production reactions
\cite{Bonutti:1996ij,Bonutti_99,Bonutti:2000bv,Camerini:2004sz,Grion:2005hu}.
The main finding was a buildup of strength with rising mass number at low
invariant masses for the $\pi^+\pi^-$ final state.
The effect was not observed for the $\pi^+\pi^+$ channel where the $\sigma$
meson cannot contribute.
A similar effect was found by the Crystal Ball collaboration at BNL.
Here, an enhancement of strength at low masses was observed for heavy nuclei 
in the $\pi^- A\rightarrow A\pi^0\pi^0$ reaction \cite{Starostin:2000cb}.
In photon induced reactions, pions can be produced in the entire volume 
of the nuclei but final state interactions suppress the contributions 
from the deep interior of the nuclei.
Final state interactions can be minimized by the use of low incident 
photon energies, giving rise to low energy pions which have much 
larger mean free paths than pions that can excite 
the $\Delta$-resonance \cite{Krusche:2004uw}.
Photoproduction of the different charge states of pions from the free proton
and the quasifree neutron has previously been studied in detail with the
DAPHNE
\cite{Braghieri:1995rf,Zabrodin:1997xd,Zabrodin:1999sq,Ahrens:2003na,Ahrens:2005ia}
and TAPS detectors
\cite{Harter:1997jq,Krusche:1999tv,Wolf:2000qt,Kleber:2000qs,Langgartner:2001sg,Kotulla:2003cx}
at MAMI-B in Mainz from threshold to the second resonance region,
and for the $\pi^0\pi^0$ channel at higher incident photon energies at
GRAAL in Grenoble \cite{Assafiri:2003mv}.
See Ref.~\cite{Krusche:2003ik} for an overview.

First results from a measurement of double $\pi^0$ and $\pi^0\pi^{\pm}$
photoproduction off carbon and lead have been reported in
\cite{Messchendorp:2002au}.
A shift of the strength to lower invariant masses was found for the heavier
nucleus for the $\pi^0\pi^0$ channel but not for the mixed charge channel.
In Ref.~\cite{Bloch:2007ka}, more experimental detail and the results of 
an additional measurement of double pion photoproduction off calcium nuclei 
are presented and compared to model calculations.
The invariant mass spectra show a similar effect as already reported in
Ref.~\cite{Messchendorp:2002au} for carbon and lead nuclei, namely a softening
of the $\pi^0\pi^0$ distributions relative to the $\pi^0\pi^{\pm}$
distributions.
The strength of the effect is comparable to that from carbon.
The data have been compared to calculations in the framework of the BUU model
\cite{Buss:2006vh,Buss_06}. A sizable part of the in-medium effects can be
explained by the model by final state interaction effects, which tend to shift
rescattered pions to smaller kinetic energies. Only for the lowest incident
photon energies a small additional downward shift of the strength to small
invariant masses for the $\pi^0\pi^0$ channel may be visible.

Decisive results will come from a recently completed experimental run
\cite{prop-pipi-nucs} with the 4$\pi$ detector combination Crystal Barrel
and TAPS at MAMI-B.
Here, superior statistics has been accumulated for carbon, calcium,
and lead targets.

\section{Proton Induced Production of Pion Pairs from Light Nuclei}

In first inclusive measurements of double pionic fusion of protons
and deuterons to $^3$He, Abashian, Booth and Crowe
\cite{Booth_60,Booth_61,Booth_63} studied the momentum spectra
of the $^3$He particles with a magnetic spectrometer.
An intriguing excess of strength close to the $\pi\pi$ threshold was found.
The phenomenon has been referred to as the ABC effect.
In Ref.~\cite{Bashkanov:2005fh}, recent exclusive measurements of the reactions
$pd\rightarrow$ $^3$He $\pi^0$, $^3$He $\pi^0\pi^0$,
and $^3$He $\pi^+\pi^-$ at $T_p=0.893$ GeV are reported.
The experiment used the WASA detection system \cite{Zabierowski:2002ah}
with a deuterium pellet target at the CELSIUS storage ring.
The beam energy is chosen to maximize the ABC effect
as observed in Ref.\cite{Banaigs:1974ee}.

\begin{figure}[htb]
\centering
\includegraphics[width=0.5\textwidth]{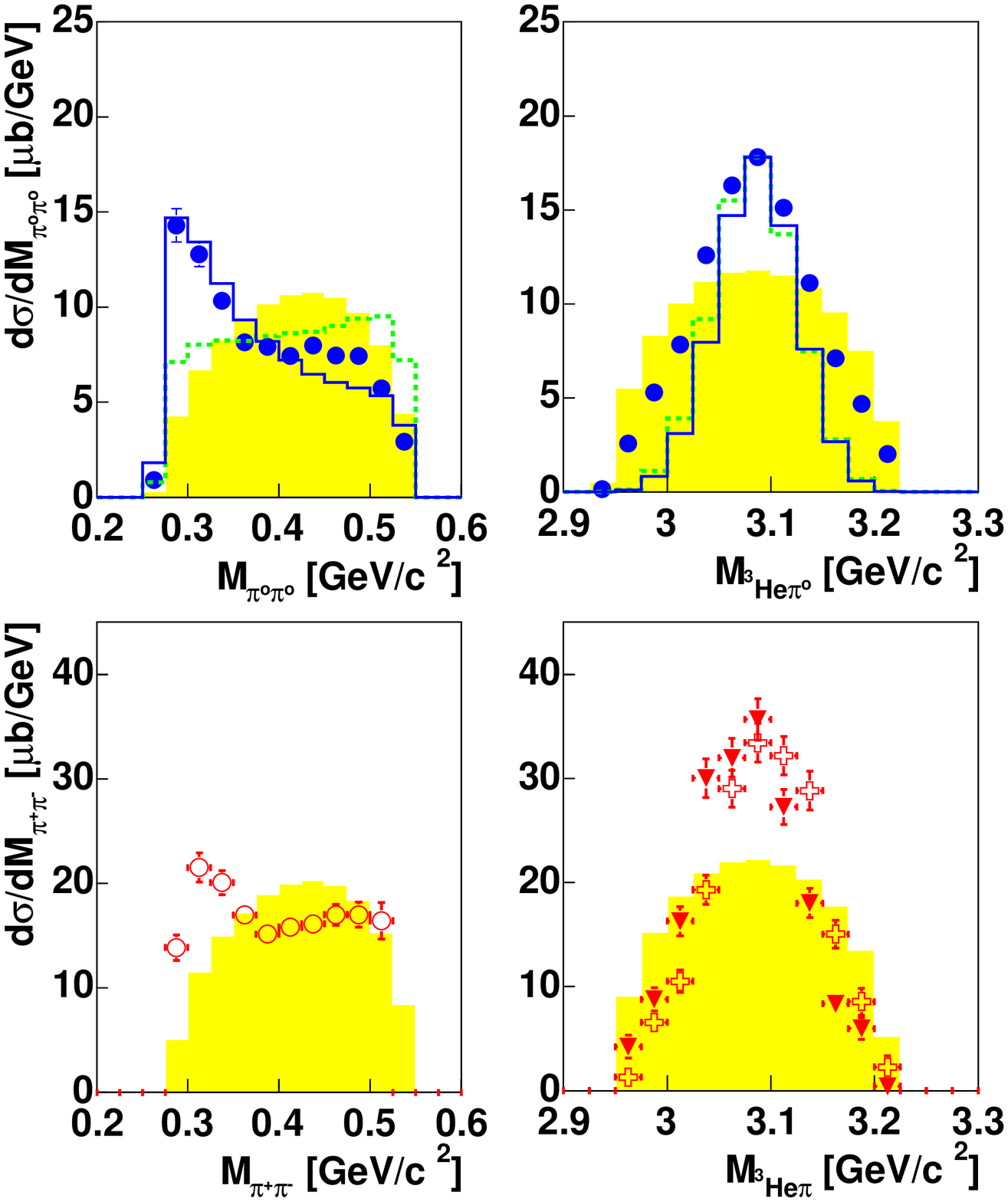}
\includegraphics[width=0.38\textwidth]{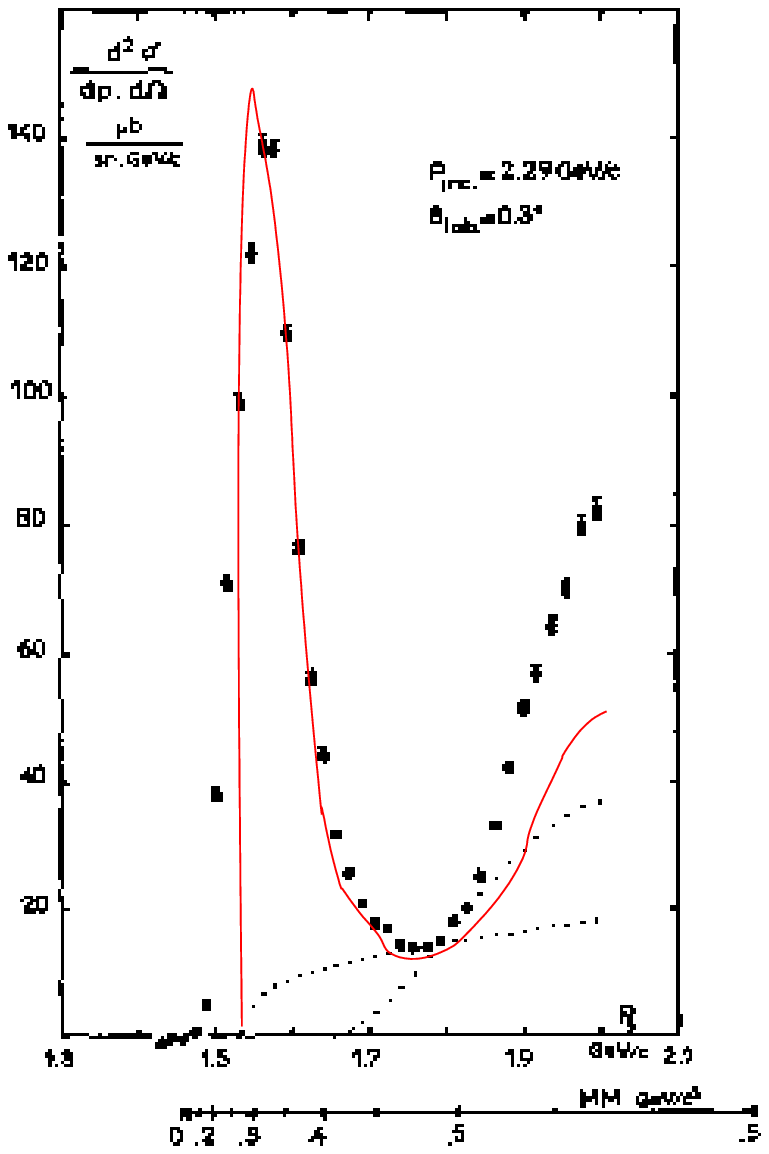}
\caption{
\underline{Left}:
Differential cross sections for the distributions of the invariant
masses $M_{\pi\pi}$ (left) and $M_{^3He \pi}$ (right, with solid
triangles for $M_{^3He \pi^-}$ and open crosses for $M_{^3{\rm He} \pi^+}$) .
Top: $pd\rightarrow$ $^3$He $\pi^0\pi^0$,
bottom: $pd\rightarrow$ $^3$He $\pi^+\pi^-$.
The shaded areas show phase space distributions for comparison.
Solid and dashed curves denote $\Delta\Delta$ calculations with
and without a bound state condition.
\underline{Right}:
Experimental $^4$He momentum spectrum for the reaction $dd \to ^4$He X
at $T_d$ = 1.09 GeV  and $\Theta_{lab}$ = 0$^\circ$ as obtained at Saclay
\cite{Banaigs:1975wt}.
The solid line displays the prediction with $\Delta\Delta$ interaction
(adjusted in height to the data) whereas the dashed lines are from the original
paper (phase space distributions from for two- and three-pion production),
adjusted to touch the data.
From Ref.~\cite{Bashkanov:2006}.
}\label{fig:1}
\end{figure}

In the left panels of Fig.~\ref{fig:1}, 
the situation is shown for the reactions
$pd\to^3$He$\pi^0\pi^0$ and $pd\to^3$He$\pi^+\pi^-$ at Tp = 0.895 GeV.
For the $\pi^0\pi^0$ channels associated with a bound deuteron
\cite{Khakimova:2007} or an unbound pp system \cite{Skorodko:2007}
in the final nuclear system the situation is similar.
The $\pi^0\pi^0$ channel is free of isospin I=1 contributions and
 in all cases exhibits a low mass enhancement.
For the $\pi^+\pi^-$ channel, which may contain isovector contributions,
the observed threshold enhancements are less pronounced in case of $^3$He
and are even absent in case of the unbound pp system in the final state.
From the angular distribution in the $\pi\pi$ subsystem it is seen that
the threshold  enhancement is of scalar nature,
see Ref.~\cite{Bashkanov:2005fh}.
The available $\Delta\Delta$ calculations fail but a missing piece is provided
by a strong $\Delta\Delta$ attraction which is able to describe the exclusive
data for d and $^3$He fusion as well as the inclusive spectra for double pionic
fusion to $^4$He, see the right panel of Fig.~\ref{fig:1}.
In summary, a strong enhancement in the low $M_{\pi\pi}$ region is observed
to be much larger in the $\pi^0\pi^0$ than in the $\pi^+\pi^-$ channel.
The results reveal the effect to be a $\sigma$ channel phenomenon
associated with the formation of a strongly attractive $\Delta\Delta$ system.
Double pion production in few nucleon systems may possibly serve as a guideline 
for the microscopic understanding of medium effects in the $\sigma$ channel and
its possible connection with chiral restoration.

In a recently approved WASA-at-COSY experiment proposal \cite{COSYPAC:2006-168}
it is planned to conduct exclusive measurements of the pionic fusion of
deuterons to $^4$He.
Whereas single $\pi^0$ production is forbidden by isospin conservation,
double pion production is isospin-allowed albeit only in the isoscalar channel.
It has been shown in inclusive measurements of this channel
that the ABC effect is largest in the $^4$He case.
The strictly isospin selective double pionic fusion to $^4$He, where the largest
effect is expected, will be the most stringent test for the understanding
of the ABC effect.
Presently, this case also constitutes the heaviest nuclear system where
exclusive measurements of double pionic fusion can be carried out experimentally.
Measurements of the energy dependence of the double pionic fusion cross section,
predicted to be resonance-like, will enable a simultaneous check the energy
dependent background for the anticipated single-pion production on $^4$He.

\section*{Acknowledgements}

The author thanks the Yukawa Institute for Theoretical Physics
at Kyoto University.
Discussions during the YKIS2006 on "New Frontiers on QCD" were very
useful to further this work.



\begin{thebibliography}{10}
\expandafter\ifx\csname url\endcsname\relax
  \def\url#1{\texttt{#1}}\fi
\expandafter\ifx\csname urlprefix\endcsname\relax\def\urlprefix{URL }\fi

\bibitem{Bernard:1987im}
V.~Bernard, U.~G. Meissner, I.~Zahed, Phys. Rev. Lett. 59 (1987) 966.

\bibitem{Yao:2006px}
W.~M. Yao, et~al., J. Phys. G33 (2006) 1--1232.

\bibitem{Caprini:2005zr}
I.~Caprini, G.~Colangelo, H.~Leutwyler, Phys. Rev. Lett. 96 (2006) 132001.

\bibitem{Hatsuda:1999kd}
T.~Hatsuda, T.~Kunihiro, H.~Shimizu, Phys. Rev. Lett. 82 (1999) 2840--2843.

\bibitem{Aouissat:1999vx}
Z.~Aouissat, Phys. Rev. C62 (2000) 012201.

\bibitem{Chiang:1997di}
H.~C. Chiang, E.~Oset, M.~J. Vicente-Vacas, Nucl. Phys. A644 (1998) 77--92.

\bibitem{Roca:2002vd}
L.~Roca, E.~Oset, M.~J. Vicente~Vacas, Phys. Lett. B541 (2002) 77--86.

\bibitem{Chanfray:2004vb}
G.~Chanfray, D.~Davesne, M.~Ericson, M.~Martini, Eur. Phys. J. A27 (2006)
  191--198.

\bibitem{Bonutti:1996ij}
F.~Bonutti, et~al., Phys. Rev. Lett. 77 (1996) 603--606.

\bibitem{Bonutti_99}
F.~Bonutti, et~al., Phys. Rev. C60 (1999) 018201.

\bibitem{Bonutti:2000bv}
F.~Bonutti, et~al., Nucl. Phys. A677 (2000) 213--240.

\bibitem{Camerini:2004sz}
P.~Camerini, et~al., Nucl. Phys. A735 (2004) 89--110.

\bibitem{Grion:2005hu}
N.~Grion, et~al., Nucl. Phys. A763 (2005) 80--89.

\bibitem{Starostin:2000cb}
A.~Starostin, et~al., Phys. Rev. Lett. 85 (2000) 5539--5542.

\bibitem{Krusche:2004uw}
B.~Krusche, et~al., Eur. Phys. J. A22 (2004) 277--291.

\bibitem{Braghieri:1995rf}
A.~Braghieri, et~al., Phys. Lett. B363 (1995) 46--50.

\bibitem{Zabrodin:1997xd}
A.~Zabrodin, et~al., Phys. Rev. C55 (1997) 1617--1620.

\bibitem{Zabrodin:1999sq}
A.~Zabrodin, et~al., Phys. Rev. C60 (1999) 055201.

\bibitem{Ahrens:2003na}
J.~Ahrens, et~al., Phys. Lett. B551 (2003) 49--55.

\bibitem{Ahrens:2005ia}
J.~Ahrens, et~al., Phys. Lett. B624 (2005) 173--180.

\bibitem{Harter:1997jq}
F.~H{\"a}rter, et~al., Phys. Lett. B401 (1997) 229--233.

\bibitem{Krusche:1999tv}
B.~Krusche, et~al., Eur. Phys. J. A6 (1999) 309--324.

\bibitem{Wolf:2000qt}
M.~Wolf, et~al., Eur. Phys. J. A9 (2000) 5--8.

\bibitem{Kleber:2000qs}
V.~Kleber, et~al., Eur. Phys. J. A9 (2000) 1--4.

\bibitem{Langgartner:2001sg}
W.~Langg{\"a}rtner, et~al., Phys. Rev. Lett. 87 (2001) 052001.

\bibitem{Kotulla:2003cx}
M.~Kotulla, et~al., Phys. Lett. B578 (2004) 63--68.

\bibitem{Assafiri:2003mv}
Y.~Assafiri, et~al., Phys. Rev. Lett. 90 (2003) 222001.

\bibitem{Krusche:2003ik}
B.~Krusche, S.~Schadmand, Prog. Part. Nucl. Phys. 51 (2003) 399--485.

\bibitem{Messchendorp:2002au}
J.~G. Messchendorp, et~al., Phys. Rev. Lett. 89 (2002) 222302.

\bibitem{Bloch:2007ka}
F.~Bloch, et~al., nucl-ex/0703037.

\bibitem{Buss:2006vh}
O.~Buss, L.~Alvarez-Ruso, P.~Muhlich, U.~Mosel, Eur. Phys. J. A29 (2006)
  189--207.

\bibitem{Buss_06}
O.~Buss, private communication.

\bibitem{prop-pipi-nucs}
S.~Schadmand, et~al., Experiment Proposal MAMI A2/3-03, Mainz, Germany.

\bibitem{Booth_60}
N.~Booth, A.~Abashian, K.~Crowe, Phys. Rev. Lett. 5 (1960) 258.

\bibitem{Booth_61}
N.~Booth, A.~Abashian, K.~Crowe, Phys. Rev. Lett. 7 (1961) 35.

\bibitem{Booth_63}
N.~Booth, A.~Abashian, K.~Crowe, Phys. Rev. 132 (1963) 2296.

\bibitem{Bashkanov:2005fh}
M.~Bashkanov, et~al., Phys. Lett. B637 (2006) 223--228.

\bibitem{Zabierowski:2002ah}
J.~Zabierowski, et~al., Phys. Scripta T99 (2002) 159--168.

\bibitem{Banaigs:1974ee}
J.~Banaigs, et~al., Nucl. Phys. B67 (1973) 1--36.

\bibitem{Khakimova:2007}
O.~Khakimova, et~al., Int. J. Mod. Phys. A22 (2007) 617.

\bibitem{Skorodko:2007}
T.~Skorodko, et~al., Int. J. Mod. Phys. A22 (2007) 509.

\bibitem{Banaigs:1975wt}
J.~Banaigs, et~al., Nucl. Phys. B105 (1976) 52.

\bibitem{Bashkanov:2006}
M.~Bashkanov, PhD Thesis, University of T{\"u}bingen.

\bibitem{COSYPAC:2006-168}
H.~Clement, V.~Hejny, Experiment Proposal COSY 168, 2006, J{\"u}lich, Germany.

\end{thebibliography}
\end{document}